\documentclass[12pt]{article}

\newcommand{\nc}{\newcommand}
\nc{\beq}{\begin{equation}}
\nc{\eeq}{\end{equation}}
\nc{\beqa}{\begin{eqnarray}}
\nc{\eeqa}{\end{eqnarray}}

\newwrite\ffile\global\newcount\figno \global\figno=1

\def\writedef#1{}

\input epsf
\def\figin{\epsfcheck\figin}\def\figins{\epsfcheck\figins}
\def\epsfcheck{\ifx\epsfbox\UnDeFiNeD
\message{(NO epsf.tex, FIGURES WILL BE IGNORED)}
\gdef\figin##1{\vskip2in}\gdef\figins##1{\hskip.5in}
\else\message{(FIGURES WILL BE INCLUDED)}%
\gdef\figin##1{##1}\gdef\figins##1{##1}\fi}

\def\figinsert{}
\def\ifig#1#2#3{\xdef#1{fig.~\the\figno}
\writedef{#1\leftbracket fig.\noexpand~\the\figno}%
\figinsert\figin{\centerline{#3}}\medskip\centerline{\vbox{\baselineskip12pt
\advance\hsize by -1truein\center\footnotesize{  Fig.~\the\figno.} #2}}
\bigskip\endinsert\global\advance\figno by1}
\def\endinsert{}

\begin{document}

\title{\large{\bf
Color Superconductivity in High Density Effective Theory
}}

\author{
Deog Ki Hong\thanks{dkhong@hyowon.cc.pusan.ac.kr} \\
Department of Physics,
Pusan National University, \\Pusan 609-735, Korea \\ \\   }

\date{March, 2000}

\maketitle

\begin{picture}(0,0)(0,0)


\end{picture}

\vspace{-24pt}

\begin{abstract}
In this talk, I discuss the recent development
in color superconductivity in terms of effective field theory.
By investigating the Cooper pair gap equations at high density,
we see that the effective theory simplifies the gap
analysis very much, especially in finding the ground state,
the precise form of the gap, and the critical temperature.
Furthermore, the effective theory enables us to estimate
the critical density for color superconductivity, which is
found to be around $230~{\rm MeV}$ in the hard-dense-loop
approximation. Finally, I briefly mention the low-lying
spectra of color superconductor at high density.
\end{abstract}

\bigskip

\begin{center}
{\it Contribution to the Proceedings of the TMU-Yale Symposium
on the Dynamics of Gauge Fields - an External Activity of APCTP,
December 13-15 1999, Tokyo, Japan.}
\end{center}

\newpage

\section{Introduction}

At high density, quarks in dense matter interact weakly with
each other and form a Fermi sea, due to asymptotic freedom.
When the energy is much less than the quark chemical potential
($E\ll\mu$), only the quarks near the Fermi surface are relevant.
The dynamics of quarks near the Fermi surface
is effectively one-dimensional, since excitations along the Fermi
surface do not cost any energy. The momentum perpendicular to the
Fermi momentum just labels the degeneracy, similarly to the
perpendicular momentum of charged particle under external
magnetic field. This dimensional reduction due to the presence of
Fermi surface makes possible for quarks to form a Cooper pair
for any arbitrary weak attraction, since the critical coupling
for the condensation in (1+1) dimensions is zero, known
as the Cooper theorem in condensed matter.

While, in the BCS theory, such attractive force for electron Cooper pair
is provided by phonons,
for dense quark matter, where phonons are absent,
the gluon exchange interaction provides the attraction, as one-gluon
exchange interaction is attractive in the color anti-triplet channel.
One therefore expects that color anti-triplet Cooper pairs will
form and quark matter is color
superconducting, which is indeed shown more than 20 years
ago~\cite{Bar_77,Frau_78,BL_84}.

Recent development in color superconductivity, started from 1998,
was spurred by recent two seminal works.
The first one is by Alford, Rajagopal, and Wilczek~\cite{ARW_99},
who convincingly argued
that for three massless flavors, the ground state of quark matter
is a so-called color-flavor locking (CFL) phase,
in which the Cooper pair takes the following form,
neglecting the small sextet component,
\begin{equation}
\left<\psi^a_{L\alpha}(\vec p)\psi^b_{L\beta}(-\vec p)\right>
=-\left<\psi^a_{R\alpha}(\vec p)\psi^b_{R\beta}(-\vec p)\right>
=\epsilon^{abI}\epsilon_{\alpha\beta I}K(p_F),
\end{equation}
where $a,b~(=1,2,3)$ denote the color indices and
$\alpha,\beta~(=1,2,3)$ denote the flavor indices.

The interesting feature of the CFL phase is that chiral symmetry is
broken and the excitations in CFL phase have  integral multiplet of
electron charge.
Though the usual quark-antiquark condensate is absent at high density,
at least at the leading order,
the chiral symmetry is spontaneously broken in the CFL phase.
The flavor indices of Cooper pairs are
locked to their color indices so that the unbroken symmetry
that leaves the Cooper pair condensate invariant is the simultaneous
rotation in the flavor and color space, breaking both color and
chiral symmetry down to their diagonal subgroup,
\begin{equation}
SU(3)_c\times SU(3)_L\times SU(3)_R\to SU(3)_{c+L+R}.
\end{equation}

The second work is done by Son~\cite{Son:1999uk}, who
showed that the Cooper pair gap in high density quark matter
is very different from the usual BCS gap,
due to the long range (color) magnetic
interaction among quarks.
By the renormalization group (RG) analysis, aided by
the analysis of the Eliashberg equation,
he found the Cooper pair
gap depends on the coupling as,
\begin{equation}
\Delta\sim {\mu\over g_s^5}\exp\left(-{3\pi^2\over \sqrt{2}g_s}\right),
\end{equation}
which was confirmed by more careful
analysis~\cite{hong98,hong99,HMSW99,
SW99,PR99,HS99}.

In this talk, I will derive the above results in terms of a
high density effective
theory derived in~\cite{hong98,hong99}. I will also calculate
the critical temperature and the critical density and mention the
mass of low-lying excitations in the CFL phase.

\section{High density effective theory}

QCD at high density has two distinct scales; one is an extrinsic scale,
$\mu$, the quark chemical potential,  and the other is the intrinsic
scale, $\Lambda_{\rm QCD}$. If the density is high enough, two scales
are well separated, $\mu\gg\Lambda_{\rm QCD}$.  To study
a low-energy physics below a scale $\Lambda$,
an effective theory approach, where
heavy modes ($\omega>\Lambda$) are separated from
light modes ($\omega<\Lambda$) systematically, has been quite
powerful.

Since we are interested in a cold dense matter where the relevant
excitations are quasi-quarks near the Fermi surface, it will be useful
to construct an effective theory that deals only
with those relevant degrees of freedom~\cite{hong98,hong99}.
A dense matter with a fixed quark number is described
by the QCD Lagrangian density with a chemical potential $\mu$,
\begin{equation}
{\cal L}_{\rm QCD}=\bar\psi
i\!\not\!D\psi-{1\over4}F_{\mu\nu}^aF^{a\mu\nu}+\mu
\bar\psi\gamma_0\psi,
\label{lag}
\end{equation}
where the covariant derivative
$D_{\mu}=\partial_{\mu}+ig_sA_{\mu}^aT^a$ and
we neglect the mass of quarks for simplicity.

At energy just below $\mu$, we may decompose the quark momentum as
\begin{equation}
p^{\mu}=\mu v^{\mu}+l^{\mu},\quad \left|l^{\mu}\right|<\mu,
\end{equation}
where $\vec v_F$ is a Fermi velocity and $v^{\mu}=(0,\vec v_F)$.
We expand the quark propagator in powers of $1/\mu$:
\begin{eqnarray}
S_F(p)={i\over (1+i\epsilon)p^0\gamma^0-\vec p\cdot \vec\gamma
+\mu\gamma^0}
=P_+{i\gamma^0\over l\cdot V+i\epsilon l^0}+
P_-{i\gamma^0\over 2\mu}+\cdots,
\label{propagator}
\end{eqnarray}
where $V^{\mu}=(1,\vec v_F)$ and
the ellipsis denote higher order terms in $1/\mu$ expansion.
In the second line of Eq.~(\ref{propagator}) we have introduced
projection operators
\begin{equation}
P_{\pm}={1\pm\vec\alpha\cdot\vec v_F\over2},
\end{equation}
where $\vec \alpha=\gamma^0\vec \gamma$. The projection operators
$P_+$ and $P_-$ project out the states near the Fermi surface and
the states in the Dirac sea, respectively.
We see that the propagating modes are the states near the Fermi
surface.

Using the techniques developed in heavy quark effective
theory~\cite{Georgi:1990um}, we Fourier-decompose the quark field as
\begin{equation}
\psi(x)=\sum_{\vec v_F}e^{i\mu\vec v_F\cdot \vec x}\psi(\vec v_F,x)
\end{equation}
where
\begin{equation}
\psi(\vec v_F,x)=\int_{\left|l^{\mu}\right|<\mu}
{d^4l\over (2\pi)^4}\psi(\vec v_F,l)e^{-il\cdot x}
\equiv\psi_+(\vec v_F,x)+\psi_-(\vec v_F,x)
\end{equation}
with $\psi_{\pm}(\vec v_F,x)=P_{\pm}\psi(\vec v_F,x)$.
The low energy effective Lagrangian that consists of the light
degrees of freedom (gluons and $\psi_+$) is obtained
by matching all one-light-particle irreducible amplitudes in
QCD with the vertex functions in the effective theory.
As shown in~\cite{hong98,hong99}, in the effective theory
the quark propagator becomes
\begin{equation}
S_F(\vec v_F;l)={1+\vec \alpha\cdot\vec v_F\over2}
{i\gamma^0\over l\cdot V+i\epsilon l^0},
\end{equation}
and in addition to the quark-gluon minimal coupling
$-i\gamma^0V^{\mu}g_s$ there is marginal
four-Fermi interaction for quarks
with opposite Fermi velocities,
\begin{eqnarray}
{\cal L}_{\rm 4f}^1&=&
{g^S_{us;tv}\over 2\mu^2}\left[
{\psi^{\dagger}_L}_t(\vec v_F,x){\psi_L}_s(\vec v_F,x)
{\psi^{\dagger}_L}_v(-\vec v_F,x){\psi_L}_u(-\vec v_F,x)
+\left(L\leftrightarrow R\right)
\right]\nonumber\\
&+&
{g^P_{us;tv}\over 2\mu^2}\left[
{\psi^{\dagger}_L}_t(\vec v_F,x){\psi_L}_s(\vec v_F,x)
{\psi^{\dagger}_R}_v(-\vec v_F,x){\psi_R}_u(-\vec v_F,x)
+\left(L\leftrightarrow R\right)
\right].
\label{four-fermi}
\end{eqnarray}

To summarize, the high density effective theory has several
interesting features: (1) In the
leading order, only $\gamma^0$ enters in the Dirac matrices.
(2) Anti-quarks are systematically decoupled. (3) There appear
marginal four-quark operators naturally. (4) It offers a systematic
high-density expansion.

\section{Cooper pair gap}
To describe the Cooper-pair gap equation, we introduce a 8-component
field, following the Nambu-Gorkov formalism,
$\Psi(\vec
v_F,x)\equiv(\psi(\vec v_F,x),\psi_c(\vec v_F,x))^T$,
where we reverted the notation $\psi$ for $\psi_+$
and introduced the charge conjugate field
$\psi_c(\vec v_F,x)=C\bar\psi^T(-\vec v_F,x)$. The
charge conjugation matrix, $C$, satisfies
$C^{-1}\gamma_{\mu}C=-\gamma_{\mu}^T$.
The inverse propagator for the Nambu-Gorkov field is
\begin{equation}
S^{-1}(\vec v_F,l)= -i\gamma_0\pmatrix{l\cdot V &
       -\Delta^{\dagger}(l_{\parallel}) \cr
-\Delta(l_{\parallel})& l\cdot\bar V\cr},
\end{equation}
where $\bar V^{\mu}=(1,-\vec v_F)$ and $\Delta$ is the Cooper-pair
gap.

The effective action for the fermion two-point function $S$ is
given as
\begin{equation}
\Gamma=-{\rm Tr}\ln S^{-1}+{\rm Tr}\left(S^{-1}-S^{-1}_0\right)S
+\left({\rm 2PI~~ diagrams}\right),
\end{equation}
where the 2PI diagrams are two-particle irreducible vacuum diagrams.
For the gluon propagator, we use an in-medium propagator,
which is in the hard dense loop (HDL) approximation given as
\begin{equation}
iD_{\mu\nu}(k)={P_{\mu\nu}^{T}\over k^2-G}+{P^L_{\mu\nu}\over k^2-F}
-\xi {k_{\mu}k_{\nu}\over k^4},
\end{equation}
where $\xi$ is the gauge parameter and
the projectors are defined by
\begin{eqnarray}
P^T_{ij}&=&\delta_{ij}-{k_ik_j\over |\vec k|^2},
\quad P_{00}^T=0=P_{0i}^T\\
P^L_{\mu\nu}&=&-g_{\mu\nu}+{k_{\mu}k_{\nu}\over k^2}-P^T_{\mu\nu}.
\end{eqnarray}
The medium effect is incorporated in $F$ and $G$, which becomes
in the weak coupling limit
($\left|k_0\right|\ll\left|\vec k\right|$)
\begin{eqnarray}
F(k_0,\vec k)\simeq M^2,\quad
G(k_0,\vec k)\simeq {\pi\over 4}M^2{k_0\over |\vec k|},
\end{eqnarray}
where $M=\sqrt{N_f/2}~g_s\mu/\pi$, the Debye mass.
The gap equations, obtained by extremizing the effective
action,  $0=\delta\Gamma/ \delta S$, are given in Euclidean space as
\begin{eqnarray}
\Delta(p_{\parallel})\!&=&\!\int{d^4q\over (2\pi)^4}
\left[-{2\over3}g_s^2
\left\{{V\cdot P^T\cdot \bar V
\over (p-q)_{\parallel}^2+
\vec q_{\perp}^2+{\pi\over4}M^2|p_0-q_0|/
|\vec p-\vec q|}\right.\right.\nonumber\\
&-&\left.\left.{1\over (p-q)_{\parallel}^2+{\vec q_{\perp}}^2+M^2}
-\xi{(p-q)_{\parallel}^2\over
(p-q)^4}\right\}+{g_{\bar3}\over \mu^2}\right]
{\Delta(q_{\parallel})\over q_{\parallel}^2+\Delta^2(q_{\parallel})},
\label{gap-hdl}
\end{eqnarray}
where $g_{\bar3}$ is a four-Fermi coupling.
Since the gluon coupling is vectorial, the gluon exchange
interaction in the gap equation does not distinguish the handedness
of quarks and thus it will generate same condensates regardless of
handedness;
$\left|\left<\psi_L\psi_L\right>\right|=
\left|\left<\psi_R\psi_R\right>\right|
=\left|\left<\psi_L\psi_R\right>\right|$, suppressing other
quantum numbers. But, the four-Fermi
interaction in the effective Lagrangian, Eq.~(\ref{four-fermi}),
lifts the degeneracy,  since
the gap in $LL$ or $RR$ channel
will be bigger than the one in $LR$ channel due to the difference in
the four-Fermi couplings, $g^S>g^P$.
The $LL$ or $RR$
condensate is energetically more preferred than the $LR$ condensate.
We also note that since in the effective theory the gluons are blind
not only to flavors but also to the Dirac indices of quarks,
the diquark Cooper-pair can be written as color anti-triplet.

Since quarks are anti-commuting, the
only possible way to form  diquark (S-wave) condensate is either in
spin-singlet or in spin-triplet:
\begin{eqnarray}
\left<{\psi_L}^a_{i\alpha}(\vec v_F,x){\psi_L}^b_{j\beta}(-\vec v_F,x)
\right>&=&-\left<{\psi_R}^a_{i\alpha}(\vec v_F,x)
{\psi_R}^b_{j\beta}(-\vec v_F,x)\right>\\
&=&\epsilon_{ij}\epsilon^{abc}
K_{\left[\alpha\beta\right]c}(p_F)+
\delta_{ij}\epsilon^{abc}
K_{\left\{\alpha\beta\right\}c}(p_F),
\end{eqnarray}
where $a,b,c=1,2,3$ are color indices,
$\alpha,\beta,\gamma=u,d,s,\cdots,N_f$ flavor indices, and
$i,j=1,2$ spinor indices. Indices in the bracket and in the
curled bracket are anti-symmetrized and symmetrized, respectively.
But, the spin-one component of the gap, $K_{\{\alpha\beta\}c}$,
vanishes algebraically, since $\psi(\vec v_F,x)=1/2\left(1+
\vec\alpha\cdot\vec v_F\right)\psi(\vec v_F,x)$ and
$(1+\vec\alpha\cdot\vec v_F)_{il}(1-\vec\alpha\cdot\vec v_F)_{lj}=0$.

When $N_f=3$, the spin-zero component of
the condensate  becomes (flavor)
anti-triplet,
\begin{equation}
K_{\left[\alpha\beta\right]c}(p_F)
=\epsilon_{\alpha\beta\gamma}K_c^{\gamma}(p_F).
\end{equation}
Using the global color and flavor symmetry, one can
always diagonalize the spin-zero condensate as
$K_c^{\gamma}=\delta_c^{\gamma}K_{\gamma}$.
To determine the parameters, $K_u$, $K_d$, and $K_s$,
we need to minimize the vacuum energy for the
condensate. The vacuum energy
is given as in the leading HDL approximation
\begin{eqnarray}
V(\Delta)\simeq
{\mu^2\over4\pi}\sum_{i=1}^9\int {d^2l_{\parallel}\over
(2\pi)^2}\left[\ln\left({l_{\parallel}^2\over
l_{\parallel}^2+\Delta_{i}^2(l_{\parallel})} \right)+
{1\over2}\cdot{\Delta_{i}^2(l_{\parallel})\over
l_{\parallel}^2+\Delta_{i}^2(l_{\parallel})}\right],
\label{ve}
\end{eqnarray}
where $\Delta_i$'s are the eigenvalues of color anti-symmetric
and flavor anti-symmetric $9\times9$ gap, $\Delta_{\alpha\beta}^{ab}$.

Approximating $\Delta_i$ to be constant,
one can easily perform the momentum integration in
(\ref{ve}) to get
\begin{eqnarray}
V(\Delta)\simeq
-0.43 {\mu^2\over 4\pi^2}
    \sum_i\left|\Delta_i(0)\right|^2.
\end{eqnarray}
Were $\Delta_i$ independent of each other, the global minimum
should occur at $\Delta_i(0)={\rm const.}$ for all $i=1,\cdots,9$.
But, due to the global
color and flavor symmetry, only three of them are independent.
Similarly to the condensate,
the gap can be also diagonalized by the color and flavor symmetry as
\begin{equation}
\Delta^{\alpha\beta}_{ab}=\epsilon_{\alpha\beta\gamma}
\epsilon^{abc}\Delta_{\gamma}\delta^{\gamma}_c.
\end{equation}
Without loss of generality, we can take $\left|\Delta_u\right|
\ge \left|\Delta_d\right|\ge\left|\Delta_s\right|$.
Let $\Delta_d/\Delta_u=x$ and $\Delta_s/\Delta_u=y$.
Then, the vacuum energy becomes
\begin{equation}
V(\Delta)\simeq -0.43 {\mu^2\over 4\pi^2}\left|\Delta_u\right|^2f(x,y),
\end{equation}
where $f(x,y)$ is a complicate function of $-1\le x,y\le 1$
that has a maximum at $x=1=y$, $f(x,y)\le 13.4$.
Therefore, the vacuum energy has a global minimum when
$\Delta_u=\Delta_d=\Delta_s$, or in terms of the eigenvalues of the gap
\begin{equation}
\Delta_i=\Delta_u \cdot(1,1,1,-1,1,-1,1,-1,-2).
\end{equation}

Now, we analyze the SD gap equation Eq.~(\ref{gap-hdl})
to see if it admits a nontrivial
solution. Since the color-flavor locking gap is
preferred if it exists, we may write the gap as
\begin{equation}
\Delta_{\alpha\beta}^{ab}=\epsilon^{abI}\epsilon_{\alpha\beta I}\Delta.
\end{equation}
We first note that the main contribution
to the integration comes from
the loop momenta in the region $q_{\parallel}^2\sim \Delta^2$ and
$|\vec q_{\perp}|\sim M^{2/3}\Delta^{1/3}$. Therefore, we find that
the leading contribution is by the first term due to the Landau-damped
magnetic gluons.
For this momentum range, we can take
$|\vec p-\vec q|\sim |\vec q_{\perp}|$ and
\begin{eqnarray}
V\cdot P^T\cdot{\bar V}=
-v_F^iv_F^j\left(\delta_{ij}-{\hat k}_i{\hat k}_j\right)
=-1+O\left({\Delta^{4/3}\over M^{4/3}}\right).
\end{eqnarray}
We also note that the term due to the four-Fermi operator is negligible,
since $g_{\bar3}\sim g_s^4$ at the matching scale $\mu$.

Neglecting $(p-q)_{\parallel}^2$
in the denominator to integrate over $\vec q_{\perp}$, we get
\begin{eqnarray}
\Delta(p_{\parallel})={g_s^2\over 9\pi}
\int{d^2q_{\parallel}\over (2\pi)^2}{\Delta(q_{\parallel})\over
q_{\parallel}^2+\Delta^2}\left[\ln\left(
{\mu^3\over {\pi\over4}M^2|p_0-q_0|}\right)
+{3\over2}\ln\left({\mu^2\over M^2}\right)+{3\over2}\xi\right].
\end{eqnarray}
We see that in this approximation $\Delta(p_{\parallel})\simeq
\Delta(p_0)$. Then, we can integrate over $\vec v_F\cdot\vec q$
to get
\begin{eqnarray}
\Delta(p_0)&=& {g_s^2\over 36\pi^2}\int_{-\mu}^{\mu}dq_0
{\Delta(q_0)\over \sqrt{q_0^2+\Delta^2}}
\ln\left({{\bar\Lambda}\over |p_0-q_0|}
\right)
\label{gapf}
\end{eqnarray}
where $\bar\Lambda=4\mu/\pi\cdot (\mu/M)^5e^{3/2\xi}$.
If we take $\Delta\simeq \Delta(0)$ for a rough estimate of the gap,
\begin{equation}
1={g_s^2\over 36\pi^2}
\left[\ln\left({\bar\Lambda\over\Delta}\right)\right]^2
\quad
{\rm or}\quad
\Delta\simeq\bar\Lambda \exp\left(-{6\pi\over g_s}\right).
\end{equation}
To take into account the energy dependence of the gap,
we convert the Schwinger-Dyson equation (\ref{gapf})
into a differential equation,
approximating the kernel as
\begin{equation}
\ln\left|p_0-q_0\right|\simeq
\ln\left[{\rm max}.\,(|p_0|,|q_0|)\right],
\end{equation}
to get
\begin{equation}
p\Delta^{\prime\prime}(p)+\Delta^{\prime}(p)+{2\alpha_s\over9\pi}
{\Delta(p)\over \sqrt{p^2+\Delta^2}}=0,
\label{diff}
\end{equation}
with boundary conditions $p\Delta^{\prime}=0$ at $p=\Delta$ and
$\Delta=0$ at $p=\bar\Delta$, where $p\equiv p_0$.
When $p\ll\Delta(p)$, the equation becomes
\begin{equation}
p\Delta^{\prime\prime}+\Delta^{\prime}+{r^2\over4}{\Delta(p)\over
|\Delta|}=0,
\end{equation}
where $r^2=2g_s^2/(9\pi^2)$ and $|\Delta|$ is the gap at $p=0$.
We find $\Delta(p)=|\Delta|J_0\left(r\sqrt{p/|\Delta|}\right)$
for $p\ll\|\Delta|$. When $p\gg \Delta$, the differential equation
(\ref{diff}) becomes
\begin{equation}
p\Delta^{\prime\prime}+\Delta^{\prime}+{r^2\over4}{\Delta\over p}=0,
\end{equation}
whose solution is
$\Delta(p)=B\sin\left[(r/2)\ln\bar\Lambda/ p\right]$.
By matching two solutions at the boundary $p=|\Delta|$ we get
\begin{equation}
B\simeq |\Delta|\quad {\rm and}\quad |\Delta|=\bar\Lambda
e^{-\pi/r}.
\end{equation}
The gap is therefore given as
at the leading order in the weak coupling expansion
\begin{equation}
\left|\Delta\right|=
c\cdot{\mu\over g_s^5}\exp\left(-{3\pi^2\over\sqrt{2}g_s}\right),
\end{equation}
where $c=2^7\pi^4N_f^{-5/2}e^{3\xi/2+1}$. This agrees with
the RG analysis done by Son~\cite{Son:1999uk} (see also~\cite{HS99})
and also with
the Schwinger-Dyson approach in full QCD~\cite{HMSW99,SW99,PR99}.
The $1/g_s$ behavior of the exponent of the gap at high density is due to
the double logarithmic divergence in the gap equation~(\ref{gap-hdl}),
similarly to the case of chiral symmetry breaking under external magnetic
fields~\cite{hong96,miransky}.

\section{Critical density and temperature}

In this section we calculate the
critical density and temperature.
First, we add the $1/\mu$ corrections to the gap equation
Eq.~(\ref{gap-hdl}) to see how the formation of
Cooper pair changes when the density decreases.
The leading
$1/\mu$ corrections to the quark-gluon interactions are
\begin{equation}
{\cal L}_1=-{1\over2\mu}\sum_{\vec v_F}\psi^{\dagger}(\vec
v_F,x)\left(\gamma_{\perp}\cdot D\right)^2\psi(\vec v_F,x)
=-\sum_{\vec v_F}\left[\psi^{\dagger}{D_{\perp}^2\over
2\mu}\psi+g_s\psi^{\dagger}{\sigma_{\mu\nu}F^{\mu\nu}\over
4\mu}\psi\right].
\end{equation}
In the leading order in the HDL approximation,
the loop correction to the vertex is
neglected and the quark-gluon vertex is shifted by the $1/\mu$
correction as
\begin{equation}
-ig_sv_F^i\mapsto -ig_sv_F^i-ig_s{l_{\perp}^i\over\mu},
\end{equation}
where $l_i$ is the momentum carried away from quarks by gluons.
We note that since the $1/\mu$ correction to the quark-gluon
vertex does not depend on the Fermi velocity of the quark, it
generates a repulsion for quark pairs.
For a constant gap approximation, $\Delta(p_{\parallel})
\simeq \Delta$, the gap equation becomes in the leading order,
as $p\to0$,
\begin{eqnarray}
1={g_s^2\over9\pi}\int{{\rm d}^2l_{\parallel}\over
(2\pi)^2}\left[
\ln\left({\bar\Lambda\over |l_0|}\right)-{3\over2}\right]
{1\over l_{\parallel}^2+\Delta^2}
={g_s^2\over 36\pi^2}\ln\left({\bar\Lambda\over \Delta}\right)
\left[\ln\left({\bar\Lambda\over\Delta}\right)-3\right].
\end{eqnarray}
When $\mu<\mu_c\simeq e^3\Delta$, the gap due
to the long-range color magnetic interaction disappears. Since the
phase transition for color superconducting phase is believed
to be of first order~\cite{phase,shuryak1}, we may assume that the gap
has the same dependence on the chemical potential $\mu$
as the leading order. Then, the critical density for the color
superconducting phase transition is given by
\begin{equation}
\mu_c=e^3\mu_c\exp\left[-{3\pi^2\over \sqrt{2}g_s(\mu_c)}\right].
\end{equation}
Therefore, if the strong interaction coupling is too strong
at the scale of the chemical potential, the gap does not form.
To form the Cooper pair gap, the strong coupling
at the scale of the chemical potential
has to be smaller than $g_s(\mu_c)=\pi^2/\sqrt{2}$.
By using the one-loop $\beta$ function
for three light flavors, $\beta(g_s)=-9/(16\pi^2)g_s^3$, and
the experimental value for the strong coupling constant,
$\alpha_s(1.73{\rm GeV})=0.32^{+0.031}_{-0.053}({\rm exp})
\pm0.016({\rm theo})$~\cite{ellis},
we get $0.13{\rm GeV}<\mu_c<0.31{\rm GeV}$, which
is about the same order as the one estimated by the instanton
induced four-Fermi interaction~\cite{shuryak1,shuryak2} or by general
effective four-Fermi interactions~\cite{phase}. But, this should
be taken as an order of magnitude,
since for such a small chemical
potential the higher order terms in $1/\mu$ expansion, which we
have neglected, are as important as the leading term.

We now consider the temperature effect on the
gap, which is quite important to understand the
heavy ion collision or the final stage of the evolution of giant
stars.  The super dense and hot quark matter will go
through a phase transition as it cools down by emitting weakly
interacting particles like neutrinos.

At finite temperature,  $T$,  the gap equation (\ref{gap-hdl})
becomes, following the imaginary-time formalism developed by
Matsubara,
\begin{eqnarray}
\Delta(\omega_{n^{\prime}})={g_s^2\over 9\pi}T
\sum_{n=-\infty}^{+\infty}
\int{{\rm d}q\over 2\pi}{\Delta(\omega_n)\over \omega_n^2+
\Delta^2(\omega_n)+q^2}\ln\left({\bar\Lambda\over \left|
\omega_{n^{\prime}}-\omega_n\right|}\right),
\end{eqnarray}
where $\omega_n=\pi T(2n+1)$ and $q\equiv \vec v_F\cdot\vec q$.
We now use the constant
(but temperature-dependent) gap approximation,
$\Delta(\omega_n)\simeq \Delta(T)$ for all $n$.
Taking $n^{\prime}=0$ and converting the logarithm into integration,
we get
\begin{eqnarray}
\Delta(T)={g_s^2\over 18\pi}T\sum_{n=-\infty}^{+\infty}
\int{{\rm d}q\over 2\pi}\int_0^{{\bar\Lambda}^2}{\rm d}x
{\Delta(T)\over \omega_n^2+\Delta^2(T)+q^2}\cdot
{1\over x+(\omega_n-\omega_0)^2}.
\end{eqnarray}
Using the contour integral,
one can in fact sum up over all $n$ to get
\begin{eqnarray}
1={g_s^2T\over36\pi^2}\int {\rm d}q\int_0^{{\bar\Lambda}^2}
{\rm d}x{1\over 2\pi i}\oint_C{{\rm d}\omega\over 1+e^{-\omega/T}}
\cdot{1\over\left(\omega^2-q^2-\Delta^2\right)\left[
(\omega_n-i\omega_0)^2+x\right]}.
\label{tgap}
\end{eqnarray}
Since the gap vanishes at the critical temperature,
$\Delta(T_C)=0$,
after performing the contour integration in Eq.~(\ref{tgap}),
we get
\begin{eqnarray}
1&=&{g_s^2\over36\pi^2}\int{\rm d}q\int_0^{{\bar\Lambda}^2}
{\rm d}x\left\{
{(\pi T_C)^2+x-q^2\over \left[
(\pi T_C)^2+x-q^2\right]^2+(2\pi T_Cq)^2}\cdot
{\tanh\left[q/(2T_C)\right]\over 2q}\right.\nonumber\\
& &\left.\quad\quad
+{(\pi T_C)^2+q^2-x\over \left[(\pi T_C)^2+q^2-x\right]^2
+(2\pi T_C)^2x}\cdot{ \coth\left[\sqrt{x}/(2T_C)\right]
\over \sqrt{2}}
\right\}.
\label{tgap1}
\end{eqnarray}
At high density $\bar\Lambda\gg T_C$,
the second term in the integral in Eq.~(\ref{tgap1}) is
negligible, compared to the first term, and
integrating over $x$, we get
\begin{eqnarray}
1&=&{g_s^2\over 36\pi^2}\int_0^{\lambda_c}{\rm d}y
{\tanh y\over y}\left[
\ln\left({\lambda_c^2\over (\pi/2)^2+y^2}\right)+
O\left({y^2\over \lambda_c^2}\right)\right]\nonumber\\
&=&{g_s^2\over 36\pi^2}\left[
\left(\ln\lambda_c\right)^2+2A\ln\lambda_c+{\rm const.}\right]
\nonumber
\end{eqnarray}
where we have introduced $y\equiv q/(2T_C)$ and
$\lambda_c\equiv\bar\Lambda/(2T_C)$ and $A$ is given as
\begin{equation}
A=\int_0^1{\rm d}y{\tanh y\over y}+\int_1^{\infty}{\rm d}y
{\tanh y-1\over y}=\ln \left({4\over\pi}\right)+\gamma.
\end{equation}
Therefore, we find, taking the Euler-Mascheroni constant
$\gamma\simeq0.577$,
\begin{equation}
T_C={e^A\over2}\Delta\simeq1.13~\Delta,
\end{equation}
which shows that the ratio between the critical temperature and
the Cooper-pair gap
in color superconductivity is same as
the BCS value,
$e^{\gamma}/\pi\simeq0.57$~\cite{PR99,brown99}.

\section{More on CFL}
As pointed out by Sch\"afer and Wilczek~\cite{sw1},
the low-lying particle spectrum of the CFL phase at high density
resembles that of low density hadron phase. Both phases have
pions and kaons,
arising from the chiral symmetry breaking. The baryons and mesons
at high density have integral multiplet of the electron charge,
the charge corresponding to the unbroken $U(1)$ gauge symmetry
at high density.
Since the diquark condensate provides additional baryon number
$B=2/3$, quarks in color superconductor have baryon number $B=1$.

To describe the dynamics of pions and kaons,
the chiral Lagrangian for the CFL phase at high density
has been constructed~\cite{hrz99,cg99} and it is shown in~\cite{hrz99}
that quarks in the CFL phase is realized as a topological soliton,
called superqualiton, as baryons in the hadron phase at low density.
Unlike the low density phase, the parameters in the
chiral Lagrangian can be calculated from the microscopic theory.
For instance,
the mass of Nambu-Goldstone bosons is found to be~\cite{hlm00}
\begin{equation}
m_{NG}^2 \sim
m_q^2\Delta\bar{\Delta}\ln(\mu^2/\Delta^2)/\mu^2,
\end{equation}
showing that mesons become massless at asymptotically large chemical
potential, as the Dirac mass term, $m_q\bar\psi_+\psi_-\simeq
(m_q^2/\mu)\psi_+^{\dagger}\psi_+$, vanishes
for infinite density. (See also~\cite{rwz99}.) This is confirmed
subsequently~\cite{Manuel:2000wm,Beane:2000ms}.
Another interesting feature of meson mass is that
the mass hierarchy is reversed~\cite{ss99}.
For instance, if $m_s>m_{u,d}$,
\begin{equation}
m_K<m_{\pi}.
\end{equation}
This inverse mass hierarchy is due to the fact that what we call
a kaon in the CFL phase is the fluctuation of Cooper-pairs in the
up and down flavor spaces,
\begin{equation}
{U_L}_{a\alpha}(x)\equiv\lim_{y\to x}{\left|x-y\right|^{\gamma_m}
\over\kappa}\epsilon^{ij}\epsilon_{abc}\epsilon_{\alpha\beta\gamma}
\psi^{b\beta}_{Li}(-\vec v_F,x)\psi^{c\gamma}_{Lj}(\vec v_F,y),
\end{equation}
where $\gamma_m$ is the anomalous dimension.

\section{Conclusion}
I have discussed the exciting recent developments in color
superconductivity in high density quark matter in terms of an
effective theory formalism. I have shown that the effective theory
calculation reproduces recent results on the Cooper pair gap,
the critical temperature, and on the ground state of high density QCD.
It not only simplifies the calculation very much but also
allows us to estimate the critical density.

I wish to thank the organizers of the TMU-Yale symposium for the
wonderful meeting. I am grateful to T. Lee, D.-P. Min, V. Miransky,
M.~Rho, I. Shovkovy,
L.~C.~R.~Wijewardhana,  and I.~Zahed for the collaborations
on the works described here and to
S.~Hsu, R.~Pisarski, D.~Rischke,  and T.~Sch\"afer
for helping me to understand this subject.
My research is supported by the academic research fund of
Ministry of Education, Republic of Korea, Project No.
BSRI-99-015-DI0114.


\end{document}